\documentclass[article,showpacs, twocolumn,nofootinbib]{revtex4-1}

\usepackage{amsmath}
\usepackage{amsfonts}
\usepackage{dsfont}
\usepackage{graphicx}
\usepackage{caption}
\usepackage{bbm}
\usepackage{bbold}
\usepackage{epstopdf}
\usepackage[font=scriptsize]{caption}

\usepackage{amsthm}

\newcommand{\trace}[1]{\ensuremath{{\rm tr}\left[{#1}\right]}}

\begin{document}
\title{Time symmetry in wave function collapse}
\date{\today}
\author{D.~J.~Bedingham}
\email{daniel.bedingham@philosophy.ox.ac.uk}
\author{O.~J.~E.~Maroney}
\email{owen.maroney@philosophy.ox.ac.uk}
\affiliation{Faculty of Philosophy, University of Oxford, OX2 6GG, United Kingdom.}

\begin{abstract}
The notion of a physical collapse of the wave function is embodied in dynamical collapse models. These involve a modification of the unitary evolution of the wave function such as to give a dynamical account of collapse. The resulting dynamics is at first sight time asymmetric for the simple reason that the wave function depends on those collapse events in the past but not those in the future. Here we show that dynamical wave function collapse models admit a general description that has no inbuilt direction of time. Given some simple constraints, we show that there exist empirically equivalent pictures of collapsing wave functions in both time directions, each satisfying the same dynamical rules. A preferred direction is singled out only by the asymmetric initial and final time constraints on the state of the Universe.
\end{abstract}

\maketitle

The claim that the fundamental laws of nature do not depend on the direction of time tacitly avoids the issue of the collapse of the wave function. According to the orthodox view of quantum theory, an observation leads to a collapse of the wave function in correspondence with the observed result. This updating procedure results in a state that is shaped by past measurement events but not future ones. A physical wave function collapse process therefore suggests an arrow of time in the fundamental dynamical laws.

Dynamical collapse models \cite{REP1,REP2} embody the idea of physical collapse of the wave function in a mathematically well defined way. They are an attempt to give the wave function a stochastic dynamics that reflects the way it evolves in orthodox treatments. The basic assumption is that the wave function is a well defined physical object and that the collapse which takes place, for example during a measurement, is a real feature of the wave function dynamics. The dynamical rules require no arbitrary separation of system and environment, and remarkably, it is possible to unify the unitary and collapse dynamics in a way which approximates the unitary or collapse behaviour in the appropriate situations.

At first sight we would expect such implementations of physical collapse of the wave function to define an arrow of time. However, here we show that for a large class of collapse models (including the well known GRW \cite{GRW} and CSL \cite{CSL1,CSL2} models), the time reversal asymmetry that they exhibit can be understood to result from asymmetric initial and final time constraints on the state of the Universe. This is the usual explanation of time-directed phenomena given time-symmetric laws. We will show that collapse models satisfying conditions specified below can be expressed mathematically in a framework with no inbuilt arrow of time. The framework can be seen as empirically equivalent pictures of forward and backward-in-time collapsing wave functions, each following equivalent dynamical rules. The choice of time direction is therefore a matter of convention. We will clarify the equivalence and show how the usual forward-in-time collapse dynamics is recovered.

It is already well known from the ABL formalism \cite{ABL} that the results of quantum measurements can be understood without appeal to a preferred direction of time. The usual way in which we use quantum theory is to pre-select special initial ensembles of states and to make predictions about the future. In the ABL formalism the ensembles involve both pre- and post-selection of states. It can then be shown that estimates for measurement outcomes in the intervening period do not make use of a direction of time. However, the ABL formalism fails to give an account of how this time symmetry is manifested in the time development of the state. Our formulation will give such an account, involving the physical collapse of the wave function.

We take a generic collapse model to have the dynamical rule that at certain prescribed times $t$, which may be randomly distributed, the Schr\"odinger state $|\Psi_t\rangle$ undergoes a spontaneous change
\begin{align}
|\Psi_{t}\rangle \rightarrow |\Psi_{t+}\rangle =  L(z_t) |\Psi_{t}\rangle,
\label{dyn}
\end{align}
where $L=L^{\dagger}$ is a Schr\"odinger picture collapse operator (see below) dependent on a random variable $z_t$. At all other times the state evolves according to the Schr\"odinger equation with time-independent Hamiltonian $H = H^{\dagger}$. The random variable $z_t$ is drawn from the probability distribution
\begin{align}
P(z_t) =  \frac{\langle\Psi_{t}|L^2(z_t)|\Psi_{t}\rangle}{\langle\Psi_{t}|\Psi_{t}\rangle}
= \frac{\langle\Psi_{t+}|\Psi_{t+}\rangle}{\langle\Psi_{t}|\Psi_{t}\rangle}.
\label{stdP}
\end{align}
In order for this to be a probability measure the collapse operator must satisfy the completeness property
\begin{align}
\int dz L^2(z) = \mathbb{1},
\label{comp}
\end{align}
where the integration measure is the appropriate one given the form of $z_t$. With this condition $L$ is mathematically equivalent to a generalised measurement operator, the variables $z_t$ to the measurement outcomes, and the probabilities $P(z_t)$ to generalised Born rule probabilities.  In the orthodox picture our access to the quantum world is via the measurement outcomes, so equivalently in collapse models our empirical access is via the collapse outcomes.  It then makes sense to treat the collapse outcomes $z_t$ as representing the real events at definite places and times in the real world \cite{BELL}.  We will take this to be the case for any acceptable dynamical collapse model.

Perhaps the most well known example of a dynamical collapse model is the GRW model \cite{GRW} which concerns a set of distinguishable particles. We have for each particle $k$ an independent Poisson distribution in time of random collapse events with collapse operators
\begin{align}
L_k(z_t)= \left(\frac{\alpha}{\pi}\right)^{3/4} e^{-\frac{\alpha}{2}(x_k-z_t)^2},
\label{GRW}
\end{align}
where $x_k$ is the position operator for particle $k$ and $z_t\in \mathbb{R}^3$. This operator has a localising effect on the state of the $k$th particle and the fixed parameter $\alpha$ sets a fundamental length scale for the localisation process. The rate of collapses per particle is fixed and can be chosen such that individual particles are rarely affected, but a bulk mass with large numbers of particles experiences frequent collapse events. In this way macroscopic pointers rapidly commit to definite readings.

Another example is the CSL model \cite{CSL1,CSL2} in which $z_t$ is a white noise field across space. The collapse operator is a tensor product of collapse operators acting at each point in space, each causing a quasi projection in the local smeared number density field (see Section IV of Ref.\cite{CSL2}).

A collapse event described by Eq.(\ref{dyn}) is formally equivalent to a generalised quantum measurement in terms of the effect that it has on the quantum state. Suppose (as with the GRW model) that the collapse effect is equivalent to a measurement of the particle location. An initial broad wave function for a single particle at time $t$ becomes a narrow wave function localised about the point $z_t$ at time $t+$ after the collapse. This transition is oriented in time, and in defining this state process we appear to have a preferred time direction in the dynamics.

\begin{figure}
        \begin{center}
        	\includegraphics[width=0.3\textwidth]{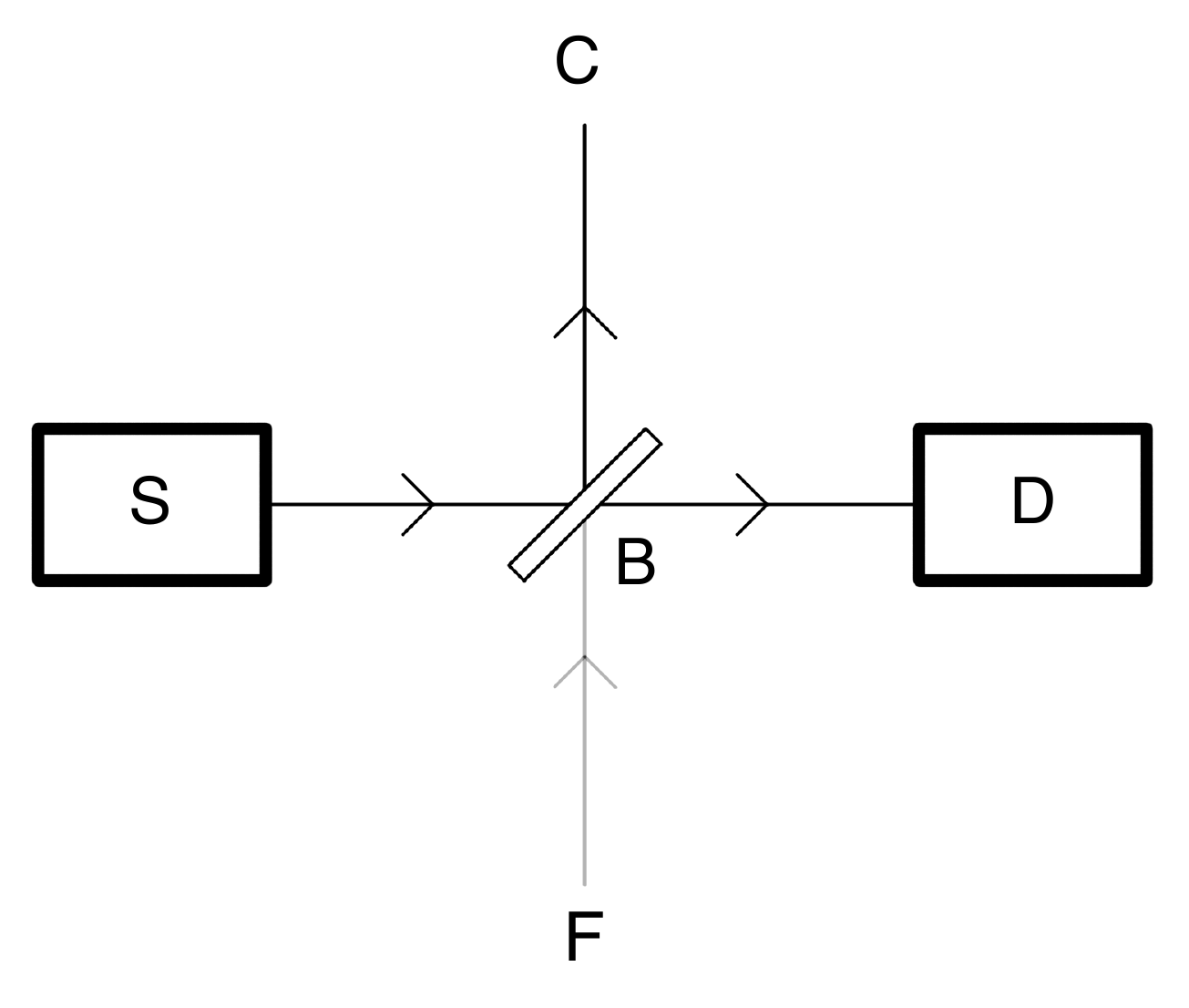}
        \end{center}
\caption{Beam splitter experiment exposing the apparent time-reversal asymmetry of wave function collapse. Particles are emitted from a source ${\sf S}$. They pass through a $50-50$ beam splitter ${\sf B}$; particles subsequently travelling to the right are detected at ${\sf D}$ and particles reflected upwards reach the ceiling ${\sf C}$. The Born rule says that half of all particles emitted from ${\sf S}$ will be detected at ${\sf D}$. Considering the events of the experiment in the backward-in-time direction we have an interaction between particle and detector at ${\sf D}$ followed with certainty by an interaction between particle and source at ${\sf S}$. This is in contrast to the Born rule prediction applied backward in time which would claim that only half of the particles travel left to ${\sf S}$ whilst half travel down to the floor ${\sf F}$.
}
\label{F1}
\end{figure}

More generally the process of wave function collapse appears to be time asymmetric without reference to a particular collapse model. This is explained with reference to a beam splitter experiment in Fig.\ref{F1} \cite{pen}.

We will now show how to construct a framework which is symmetric under time reversal, within which the standard collapse model features (\ref{dyn}) and (\ref{comp}) hold, and (\ref{stdP}) exists as a special case.  Suppose that we have an initial state of the Universe $\rho_I$ at time $t_{0}$. We take this to  mean that an initial pure state of the Universe is drawn from a distribution defined by $\rho_I$.  The collapse events occur at discrete times between the initial time $t_0$ and a final time $t_n$. These times are randomly generated, with a probability distribution that is independent of the quantum state and has no intrinsic time asymmetry. Continuous collapse processes are a limit case of this construction \cite{CSL2}. The set of collapse times are given: $\{t_i\} = \{t_1,t_2,\ldots,t_{n-1}\}$, such that $t_{0}<t_{1}<t_{2}<\cdots<t_n$. We label the collapse outcome at time $t_i$ by $z_i$.

Using the forward-in-time state dynamics (\ref{dyn},\ref{stdP}), the probability for collapse events occurring up to some time $t$ is
\begin{align}
P(\{ z_i | t_i < t\}|\rho_{I} ) = {\rm tr}[\pi_t],
\label{ForwardProbUncon}
\end{align}
where, given that $t_j < t \leq t_{j+1}$, the operator
\begin{align}
\pi_{t} =  U(t&-t_j)L({z}_{j}) U_{j,j-1} \cdots U_{2,1}L({z}_{1}){U}_{1,0}\rho_I
\nonumber \\
& {U}_{0,1}L({z}_{1})U_{1,2} \cdots U_{j-1,j}L({z}_{j})U(t_j-t),
\label{history}
\end{align}
with $U_{i,k} =U(t_i-t_k) =  e^{-iH(t_i-t_k)}$. From this we can identify
\begin{align}
\rho_{t} = \frac{\pi_{t}} {\trace{\pi_{t}}},
\label{rhot}
\end{align}
as the quantum state at $t$ that results from the collapse events $\{z_i| t_i < t\}$. We will use the notation $\rho_i = \rho_{t_i}$ and $\pi_i = \pi_{t_i}$.

Now suppose that the final state of the Universe at time $t_n$ is conditioned on the mixed state $\rho_F$, understood as being equivalent to the POVM measurement outcome $\rho_F$ given the preceding dynamics. A complete set of POVM elements is given by $\{\rho_F,\mathbb{1}-\rho_F\}$: the probability of the outcome $(\rho_F \text{ is True})$ for a state $\rho$ is given by the standard formula ${\rm tr}\left[\rho_F\rho\right] $.

The probability for the complete set of collapse events occurring in the duration of the entire Universe, $\{z_i\} = \{z_1,z_2,\ldots,z_{n-1}\}$, conditioned on the initial and final states is given, using Bayes' theorem, by
\begin{align}
P(\{ z_i\}|\rho_{I} , \rho_{F}) =\frac{P(\rho_{F} | \{ z_i\} , \rho_{I})P(\{z_i\}|\rho_I)}{P(\rho_{F} | \rho_{I})}.
\label{probmix0}
\end{align}
We can use (\ref{rhot}) to find $P(\rho_{F} | \{ z_i\} , \rho_{I}) = \trace{\rho_{F}\rho_{n}}$; from (\ref{ForwardProbUncon}) we have $P(\{z_i\}|\rho_I) = \trace{\pi_{n}}$; and averaging over collapse outcomes gives
\begin{align}
P(\rho_{F} | \rho_{I})  = \int dz_1\cdots dz_{n-1} P(\rho_{F} | \{ z_i\} , \rho_{I})P(\{z_i\}|\rho_I).
\end{align}
We then have
\begin{align}
P(\{ z_i\}&|\rho_{I} , \rho_{F}) =\frac{{\rm tr}\left[\rho_F \pi_{n} \right]}
{\int dz_1\cdots dz_{n-1}{\rm tr}\left[\rho_F \pi_{n} \right]}.
\label{probmix}
\end{align}

We now show that the structure of Eq.(\ref{probmix}) is time symmetric if there exists a complete set of basis states $\{|\phi_i\rangle \}$ such that
\begin{align}
\langle \phi_i | U (t) |\phi_j\rangle^* &=   \langle \phi_i | U (-t) |\phi_j\rangle;
\nonumber\\
\langle \phi_i | L(z) |\phi_j\rangle^* &=   \langle \phi_i | L(z) |\phi_j\rangle.
\label{conj}
\end{align}
This means that there exists a basis in which both $U$ and $L$ are symmetric matrices.

First we introduce a new time coordinate $\bar{t}$. This is related to the original coordinate by $\bar{t} = -t$. We also introduce a new labelling for the set of events $\{\bar{0},\bar{1},\bar{2},\ldots,\bar{n}\}$. These are related to the original labels $\{0,1,2,\ldots,n\}$ by $\bar{j} \equiv n-j$ meaning that the event $\bar{j}$ in the new labelling system and the event $n-j$ in the original labelling system are the same event. In terms of the new time coordinate, the time at which the event $\bar{j}$ occurs is written $\bar{t}_j = -t_{n-j}$ (this event occurs at time $t_{n-j}$ using the original time coordinate). We also relabel the collapse outcome associated with this event by $\bar{z}_j=z_{n-j}$. This is summarised in Table~\ref{tab:T1}.

\begin{table}[t]
\caption{Forward-in-time and backward-in-time notation}
\centering
\begin{tabular}{ccccccc}
\hline \hline
\multicolumn{2}{c}{Forward-in-time} &&& \multicolumn{2}{c}{Backward-in-time} &\\
\hline
 Event & Time & Outcome && Event & Time & Outcome \\
\hline
0 & $t_0$ &  $\rho_I$ &&  $\bar{n}$ &$\bar{t}_n = -t_0$ &  $\rho_I^*$\\
1 & $t_1$ & $z_1$ && $\overline{n-1}$&$\bar{t}_{n-1} = -t_1$ & $\bar{z}_{n-1} = z_1$ \\
2 & $t_2$ & $z_2$ && $\overline{n-2}$ &$\bar{t}_{n-2} = -t_2$ & $\bar{z}_{n-2} = z_2$ \\
$\vdots$ & $\vdots$ & $\vdots$ && \vdots &$\vdots$ & $\vdots$ \\
$n-1$ & $t_{n-1}$ & $z_{n-1}$ && $\bar{1}$ &$\bar{t}_{1} = -t_{n-1}$ & $\bar{z}_{1} = z_{n-1}$ \\
$n$ & $t_n$ & $\rho_F$ && $\bar{0}$ &$\bar{t}_{0} = -t_n$ & $\rho_F^*$\\
\hline
\end{tabular}
\label{tab:T1}
\end{table}

Under the new coordinate system the ordering of events is reversed yet $\bar{t}_0 < \bar{t}_1 < \bar{t}_2<\cdots <\bar{t}_n$. If the state dynamics operating in the time direction defined by $\bar{t}$, make the same prediction for the probability of the collapse outcomes $\{\bar{z}_i\}$ as that given by Eq.~(\ref{probmix}), then the dynamics are structurally time symmetric. This means that for a given set of collapse outcomes, the time reversed sequence exists as a solution to the state dynamics with the same probability (assuming appropriately reversed boundary conditions), and so, given a complete set of collapse outcomes, there is no way to distinguish a preferred direction of time in the dynamical scheme by which collapse events are generated.

To show this we define $A^*$ by
\begin{align}
\langle \phi_i |A^*|\phi_j \rangle = \langle \phi_i |A|\phi_j \rangle^*,
\end{align}
and note that $(AB)^*=A^* B^*$. Since $\trace{\rho_F\pi_n}$ is real we have
\begin{align}
\trace{\rho_F\pi_n} = \trace{\pi_n^*\rho_F^*}.
\label{trrel1}
\end{align}
Using (\ref{history}) and (\ref{conj}) we find
\begin{align}
 \pi^*_n  &= (U_{n,n-1}L(z_{n-1})\cdots U_{2,1}L(z_1)U_{1,0}\rho_I
\nonumber\\ &\quad\quad\quad U_{0,1} L(z_1)U_{1,2}\cdots L(z_{n-1})U_{n-1,n} )^*
\nonumber\\
& = U^*_{n,n-1}L^*(z_{n-1})\cdots U^*_{2,1}L^*(z_1)U_{1,0}\rho^*_I
\nonumber\\ &\quad\quad\quad U^*_{0,1} L^*(z_1)U^*_{1,2}\cdots L^*(z_{n-1})U^*_{n-1,n} )
\nonumber\\
& = U_{n-1,n}L(z_{n-1})\cdots U_{1,2}L(z_1)U_{0,1}\rho^*_I
\nonumber\\ &\quad\quad\quad U_{1,0} L(z_1)U_{2,1}\cdots L(z_{n-1})U_{n,n-1} .
\end{align}
We convert to the new coordinate system using the correspondences outlined above and the notation $\bar{U}_{i,j}=U(\bar{t_i}-\bar{t_j})=U(t_{n-j}-t_{n-i})=U_{n-j,n-i}$,
\begin{align}
\pi^*_n =\bar{U}_{0,1}&L(\bar{z}_1)\cdots \bar{U}_{n-2,n-1}L(\bar{z}_{n-1})\bar{U}_{n-1,n}\rho^*_I
\nonumber\\
&\bar{U}_{n,n-1} L(\bar{z}_{n-1})\bar{U}_{n-1,n-2}\cdots L(\bar{z}_1)\bar{U}_{1,0}.
\end{align}
We use the cyclic property of the trace to show
\begin{align}
\trace{\pi_n^*\rho_F^*} = \trace{ \rho_I^* \bar{\pi}_n},
\label{trrel2}
\end{align}
where we define
\begin{align}
\bar{\pi}_{\bar{t}} = \bar{U}(\bar{t}-\bar{t}_{j})L(\bar{z}_{j})& \cdots L(\bar{z}_{1})\bar{U}_{1,0}\rho_F^*
 \nonumber\\ &\bar{U}_{0,1}L(\bar{z}_{1}) \cdots L(\bar{z}_{j})\bar{U}(\bar{t}_j-\bar{t}),
\label{rstate}
\end{align}
for $\bar{t}_j<\bar{t}\leq \bar{t}_{j+1}$, and the shorthand $\bar{\pi}_j = \bar{\pi}_{\bar{t}_j}$. Finally, from (\ref{probmix}), (\ref{trrel1}), and (\ref{trrel2}), we have
\begin{align}
P(\{ z_i\}|\rho_{I} , \rho_{F}) &=\frac{{\rm tr}\left[\rho_I^* \bar{\pi}_n\right]}
{\int d\bar{z}_1\cdots d\bar{z}_{n-1}{\rm tr}\left[\rho_I^* \bar{\pi}_n\right]}.
\end{align}
By comparison with (\ref{probmix}) we see that this formula corresponds to the probability of a sequence of collapse outcomes $\{\bar{z}_1,\bar{z}_2,\ldots,\bar{z}_{n-1}\}$ at times $\{\bar{t}_1,\bar{t}_2,\ldots,\bar{t}_{n-1}\}$ (the reverse sequence) given the initial condition $\rho^*_F$ at $\bar{t}_0$ and the final constraint $\rho_I^*$ at $\bar{t}_n$ (reversed boundary conditions). This means that the original sequence of collapse outcomes, $\{z_1,z_2,\ldots,z_{n-1}\}$ at times $\{t_1,t_2,\ldots,t_{n-1}\}$ given the initial condition $\rho_I$ at $t_0$ and the final constraint $\rho_F$ at $t_n$, and the reverse sequence with reversed boundary conditions, have the same probability according to the collapse dynamics.

This implies that we can define a backward-in-time dynamics where, if there is no collapse event between $\bar{s}$ and $\bar{t}>\bar{s}$, then
\begin{align}
|\bar{\Psi}_{\bar{t}}\rangle = U(\bar{t}-\bar{s})|\bar{\Psi}_{\bar{s}}\rangle,
\end{align}
and if there is a collapse event at time $\bar{t}$, then
\begin{align}
|\bar{\Psi}_{\bar{t}}\rangle \rightarrow |\bar{\Psi}_{\bar{t}+}\rangle =  L(\bar{z}_{\bar{t}}) |\bar{\Psi}_{\bar{t}}\rangle.
\end{align}
The backward-in-time dynamics is identical in form to the forward-in-time dynamics. Each provide a valid basis for understanding how the complete set of collapse outcomes are generated via the stochastic development of the state. Given the future boundary condition we can determine the backward-in-time state at time $\bar{t}$ which, using (\ref{rstate}), is $\bar{\rho}_t = \bar{\pi}_t/\trace{\bar{\pi}_t}$. At time $\bar{t}_j$ we have $\bar{\rho}_j = \bar{\rho}_{\bar{t}_j}$. 

However, for a given $\{z_i\}$ there is no reason to expect that this backward-in-time state is the same as the forward-in-time state $\rho_{n-j}$ (at time $t_{n-j}$ in the original time coordinate). The forward going state $\rho_{n-j}$, results from the past boundary condition and all collapses in the past, and makes predictions about future collapses, whereas the backward going state, $\bar{\rho}_{j}$, results from the future boundary condition and all collapses in the future, and makes retrodictions about past collapses.  If these two states were empirically distinguishable, one could identify the correct one by observation. However, since the collapse outcomes form an empirically adequate description of the quantum world \cite{BELL}, then there exists both an underlying forward-in-time or backward-in-time state dynamics, each consistent with observations. This implies that the formulation has no intrinsic arrow of time. The only possible remaining sources of time asymmetry are asymmetric boundary conditions of the Universe.

We note that for the GRW model where $H = \sum_k p_k^2/2m + V(\{x_k\})$, and the collapse operator is given by (\ref{GRW}), we can choose the improper particle position state basis to show that the conditions (\ref{conj}) are satisfied. For the CSL model we can use the improper vectors $a^{\dagger}(x_1)\cdots a^{\dagger}(x_k)|0\rangle$, where $a^{\dagger}(x)$ is the particle creation operator at point $x$, to show the same. Both the GRW and CSL models can therefore be understood in the time-symmetric framework.

We now show how the usual Born rule probabilities for individual collapse events (\ref{stdP}) are recovered through the asymmetric use of boundary conditions. Suppose that the present time is $t_j$ and we condition on all the past collapse outcomes $\{z_i |i<j\}$. Together with $\rho_I$ this is equivalent to conditioning on the state at time $t_j$. The probability for all collapses at or later than $t_j$, $\{z_i | i\geq j \}$ is then given by
\begin{align}
P(\{ z_i|  i\geq & j \}|\rho_I , \rho_F , \{ z_i| i<j\})\nonumber\\
&=\frac{P(\{ z_i\},\rho_F|\rho_I  )}{P(\{ z_i|  i< j \},\rho_F|\rho_I )}\nonumber\\
&= \frac{{\rm tr}\left[\bar{\rho}^*_{n-j}L(z_{j}) \rho_j L(z_{j})\right]}
{\int dz_j\cdots dz_{n-1}{\rm tr}\left[\bar{\rho}^*_{n-j}L(z_{j}) \rho_j L(z_{j})\right]}.
\label{single0}
\end{align}
The probability for the collapse event at $t_j$ alone, conditional on past collapse outcomes is
\begin{align}
P(z_j & |\rho_I , \rho_F , \{ z_i| i<j\}) \nonumber\\
&= \int dz_{j+1}\cdots dz_{n-1} P(\{ z_i| i\geq j\}|\rho_I , \rho_F , \{ z_i|i<j \})\nonumber\\
&=\frac{\int dz_{j+1}\cdots dz_{n-1}{\rm tr}\left[\bar{\rho}^*_{n-j}L(z_{j}) \rho_j L(z_{j})\right]}
{\int dz_j\cdots dz_{n-1}{\rm tr}\left[\bar{\rho}^*_{n-j}L(z_{j}) \rho_j L(z_{j})\right]}.
\label{genP}
\end{align}
Now, if
\begin{align}
\int d\bar{z}_1\cdots d\bar{z}_{n-j-1} \bar{\rho}_{n-j}\propto \mathbb{1},
\label{unif}
\end{align}
then the formula (\ref{genP}) reduces to
\begin{align}
P(z_j |\rho_I , \rho_F , \{ z_i| i<j\}) &= \frac{ {\rm tr}\left[L(z_j)^2\rho_j\right]}
{{\rm tr}\left[\rho_j\right]} .
\label{genbornrec}
\end{align}
This is the standard Born rule probability for a forward-in-time collapse model with a mixed state $\rho_j$ at time $t_j$. The condition (\ref{unif}) effectively means that the final state of the Universe is shielded by sufficiently many future collapse events that it has no present influence. Put more precisely, in an hypothetical ensemble of Universes evolved from the final time using the backward-in-time dynamics to time $\bar{t}_{n-j} = -t_j$ the collapse dynamics have dispersed the final state uniformly across all basis states. We can further weaken the constraint (\ref{unif}) since all we need to recover (\ref{genbornrec}) is for $\int d\bar{z}_1\cdots d\bar{z}_{n-j-1}\bar{\rho}_{n-j}$ to be approximately uniform over the space of possible states $L(z_j)\rho_jL(z_j)$ given possible random values of $z_j$.

We note that (\ref{unif}) is trivially satisfied in the special case where $\rho_F \propto \mathbb{1}$. Further choosing the initial state to be a pure state $\rho_I =  |\Psi_I\rangle\langle\Psi_I|$, we recover (\ref{stdP}) precisely:
\begin{align}
P(z_j |\rho_I =  |\Psi_I\rangle\langle\Psi_I|, \rho_F \propto \mathbb{1}  , &\{ z_i| i<j\})
 \nonumber\\
&=\frac{ \langle\Psi_{j}|L(z_j)^2|\Psi_j\rangle}
{\langle\Psi_{j}|\Psi_j\rangle},
\end{align}
where
\begin{align}
|\Psi_j\rangle = {U}_{j,j-1}L({z}_{j-1}) \cdots L({z}_{1}){U}_{1,0}|\Psi_I\rangle.
\end{align}

Due to the structural time symmetry of the probability rule (\ref{probmix}) we can also construct the equivalent backward-in-time formula where we condition on all future collapses $\{\bar{z}_i|i<j\}$=$\{{z}_i|i>n-j\}$. Together with $\rho_F$, this is equivalent to conditioning on the backward-in-time state at time $\bar{t}_j$, $\bar{\rho}_j$,
\begin{align}
P(\bar{z}_j |\rho_I , &\rho_F , \{ \bar{z}_i| i<j\}) =\nonumber\\
&\frac{\int d\bar{z}_{j+1}\cdots d\bar{z}_{n-1}{\rm tr}\left[{\rho}^*_{n-j}L(\bar{z}_{j}) \bar{\rho}_{j} L(\bar{z}_{j})\right]}
{\int d\bar{z}_{j}\cdots d\bar{z}_{n-1}{\rm tr}\left[{\rho}^*_{n-j} L(\bar{z}_{j}) \bar{\rho}_{j}  L(\bar{z}_{j})\right]}.
\label{single}
\end{align}
In the special cases where
\begin{align}
\int dz_1\cdots dz_{n-j-1} \rho_{n-j} \propto \mathbb{1},
\label{cont}
\end{align}
we find
\begin{align}
P(\bar{z}_{j} |\rho_I , \rho_F , \{ \bar{z}_{i}| i<j\}) =
\frac{ {\rm tr}\left[L(\bar{z}_{j})^2\bar{\rho}_{j}\right]}
{{\rm tr}\left[\bar{\rho}_{j}\right]} ,
\label{bornback}
\end{align}
so that the Born rule applies backwards in time. Here (\ref{cont}) would require that the initial condition is shielded by sufficiently many past collapse events as to have no present effect: in an hypothetical ensemble of Universes evolved to time $t_{n-j}$ the collapse dynamics have dispersed the initial state uniformly across all basis states.

In general, if (\ref{cont}) does not hold then (\ref{single}) signals a breakdown in the Born rule for the backward-in-time dynamics as the collapse probabilities are influenced by the presence of the initial condition. This feature accounts for the observed wave function collapse asymmetry described in Fig.\ref{F1}. The asymmetry results from the special initial state of the Universe, which enables the pre-selection of special initial ensembles of states for experiments.

In the original dynamical collapse picture we had a quantum state evolving forward-in-time and some set of collapse outcomes generated by the dynamics. In the time-symmetric formulation the only thing that we have added is a final time boundary condition. Indeed by choosing this final boundary condition to be the uniformly mixed state we recover the original collapse dynamics. What the final boundary condition does is to allow us to construct a backward-in-time collapsing quantum state, a counterpart to the usual forward-in-time collapsing quantum state. At any given point in time there are then two quantum states---one that carries information about the past and one that carries information about the future. What ties these two pictures together is a consistent set of collapse outcomes $\{z_i\}$, assumed to provide an empirically adequate description of the quantum world.

We have demonstrated that collapse models can be understood in a framework which has no inbuilt arrow of time. More generally, this undermines the notion that physical collapse of the wave function is an inherently time-asymmetric process. Provided the conditions (\ref{conj}) hold, the claim that the fundamental dynamical laws of nature are time symmetric is therefore valid even if we assume a physical collapse of the wave function.

\end{document}